\documentclass[a4paper,11pt]{article}
\pdfoutput=1
\usepackage{jcappub,amssymb,graphicx}
\title{\boldmath Chaos in a tunneling universe}

\author{Martin Bojowald,}\emailAdd{bojowald@psu.edu}

\author{Ari Gluckman}\emailAdd{ayg5136@psu.edu} 

\affiliation{Department of Physics,
The Pennsylvania State University,\\
104 Davey Lab, University Park, PA 16802, USA}

\abstract{A recent quasiclassical description of a tunneling universe model is
shown to exhibit chaotic dynamics by an analysis of fractal dimensions in the
plane of initial values. This result relies on non-adiabatic features of the
quantum dynamics, captured by new quasiclassical methods. Chaotic dynamics in
the early universe, described by such models, implies that a larger set of
initial values of any large expanding branch can be probed.}


\begin{document}
\maketitle

\section{Introduction}

Quantum effects such as tunneling may be relevant in the early universe at
high temperature and curvature. Early models \cite{tunneling} have recently
been extended to oscillating versions
\cite{OscillatingFriedmann,OscillatingSimple} in which the evolving scale
factor, classically trapped in a finite region, may escape by quantum
tunneling and approach a singularity
\cite{OscillatingTunnel,OscillatingCollapse,OscillatingRate}. An application
\cite{TunnelingUniverse} of quasiclassical methods revealed non-trivial
features of the tunneling dynamics that is not captured by the traditional
derivation of tunneling coefficients from stationary states. In particular,
some of the dynamics was found to depend sensitively on the choice of initial
values in the trapped region, suggesting chaotic behavior. The purpose of the
present paper is to confirm this suspicion by a dedicated analysis.

The physical relevance of chaos in universe models can be seen by going beyond
the first approximation of an exactly homogeneous universe. At low curvature,
observations of large-scale structure indicate that approximate spatial
homogeneity is a good late-time assumption, but it is unlikely to hold in the
early universe. At large density and curvature, the gravitational dynamics is
rather dominated by attraction to denser regions and their subsequent
collapse, suggesting a very inhomogeneous distribution out of which our
universe may have arisen by cosmic inflation. The corresponding rapid
expansion would then have magnified and diluted a small region that eventually
formed our visible universe. In the initial distribution, however, this small
region would have been only one tiny patch. Thanks to its smallness, it may be
assumed to be nearly homogeneous and approximately described by simple
(classical or quantized) Friedmann dynamics. But its properties were
determined by high-density features that are more involved than those tested
at late times.

It has been known for some time that the classical dynamics of such a patch is
chaotic \cite{ChaosGR,Farey,ChaosRel,Billiards}, provided it includes effects
of anisotropy (while still being spatially homogeneous). Such a dynamics is
expected asymptotically close to a spacelike singularity according to the
Belinskii--Khalatnikov--Lifshitz (BKL) scenario \cite{BKL}. Given the
asymptotic nature of this model, the effects of this kind of chaos are most
pronounced in backward evolution closer and closer to the big-bang
singularity. They are therefore relevant for a conceptual analysis of possible
initial conditions at the very beginning of the universe, but their
implications for potential observations, seen for instance through the
magnifying glass of inflation, would be rather indirect.

Once certain matter effects start being relevant, the anisotropic asymptotic
geometry may isotropize \cite{Isotropize}, a property that is also desirable
for models of inflation. In an intermediate phase between an asymptotically
early BKL regime and the (still early) beginning of inflation, anisotropy may
be ignored while quantum effects are strong.  The results presented here show
that even the isotropic dynamics is chaotic if it is described
quasiclassically by including quantum fluctuation terms, applied in the
specific analysis to tunneling-type potentials as in oscillating models. We
will use the same model and quasiclassical extensions as derived in
\cite{TunnelingUniverse}, reviewed in the next section, and show proofs of
chaos based on a numerical analysis of the fractal domension in a space of
initial values. In our conclusions we will demonstrate which features of the
specific potential are likely to be responsible for chaos. Compared with
BKL-type chaos, the new chaotic features identified here, closer to the onset
of inflation, may have phenomenological implications which we leave for future
analysis.

\section{Quasiclassical  model}

The classical model and its potential, introduced in
\cite{OscillatingSimple} follow from the Friedmann equation
\begin{equation} \label{Friedmann}
  \frac{\dot{a}^2}{a^2} + \frac{k}{a^2}= \frac{8\pi
    G}{3}\left(\Lambda+\frac{\sigma}{a}+\frac{p_{\phi}^2}{2a^6}\right)
\end{equation}
with positive spatial curvature, $k>0$, a negative cosmological constant
$\Lambda<0$, and two matter contributions, one with energy density $\sigma/a$
where $\sigma>0$, and one from a free, massless scalar field $\phi$ with
momentum $p_{\phi}$. Our results do not depend much on the specific features of
the contributions from $\sigma$ and $p_{\phi}$, other than the trapped
potential region they form together with the curvature term. For the latter,
we choose $k=1$, but smaller values are also possible; see
\cite{TunnelingUniverse,Infrared} for more details.

The quasiclassical methods we use are canonical. We therefore replace the time
derivative $\dot{a}$ of the scale factor with the standard momentum
\begin{equation}
 p_a=-\frac{3}{4\pi G} a \dot{a}
\end{equation}
in Friedmann cosmology; see for instance \cite{Foundations}. The Friedmann
equation (\ref{Friedmann}) can then be written as
\begin{equation} 
 0= \frac{16}{9}\pi^2G^2 p_a^2+ a^2 U_{\rm harmonic}(a)- \frac{\tilde{p}^2}{a^2}
\end{equation}
with a harmonic potential
\begin{equation} \label{Uharmonic}
  U_{\rm harmonic}(a)= \omega^2(a-\gamma/\omega)^2+k-\gamma^2
\end{equation}
expressed in terms of the parameters
\begin{equation} \label{omegagamma}
  \omega=\sqrt{-\frac{8\pi G\Lambda}{3}} \quad\mbox{and}\quad
  \gamma=\sqrt{-\frac{2\pi G\sigma^2}{3\Lambda}}\,.
\end{equation}
The scalar contribution is not harmonic, and only slightly rewritten by introducing
\begin{equation}
  \tilde{p}=\sqrt{\frac{4\pi G}{3}} p_{\phi}\,.
\end{equation}

Finally, we perform a canonical transformation from $(a,p_a)$ to
$(\alpha,p_{\alpha})$ where
\begin{equation} \label{alpha}
  \alpha=\ln(\omega\gamma a)
\end{equation}
and
\begin{equation}
  p_{\alpha}=ap_a= -\frac{3}{4\pi G} a^2\dot{a}\,.
\end{equation}
In these variables, the Friedmann equation is equivalent to
\begin{equation} \label{constraint}
  0=p_{\alpha}^2+U_p(\alpha)
\end{equation}
with the potential
\begin{equation} \label{Up}
  U_p(\alpha) = \frac{e^{4\alpha}}{\beta^2} 
  \left(k-2e^{\alpha}+\frac{e^{2\alpha}}{\gamma^2} \right) -p^2
\end{equation}
and
\begin{equation}
  \beta=\frac{4\pi G}{3} \omega^2\gamma^2=\left(\frac{4\pi G}{3}\right)^3 \sigma^2 \quad,\quad
  p=\frac{3}{4\pi G} \tilde{p}=\sqrt{\frac{3}{4\pi G}} p_{\phi}\,.
\end{equation}

\begin{figure}
\begin{center}
  \includegraphics[width=13cm]{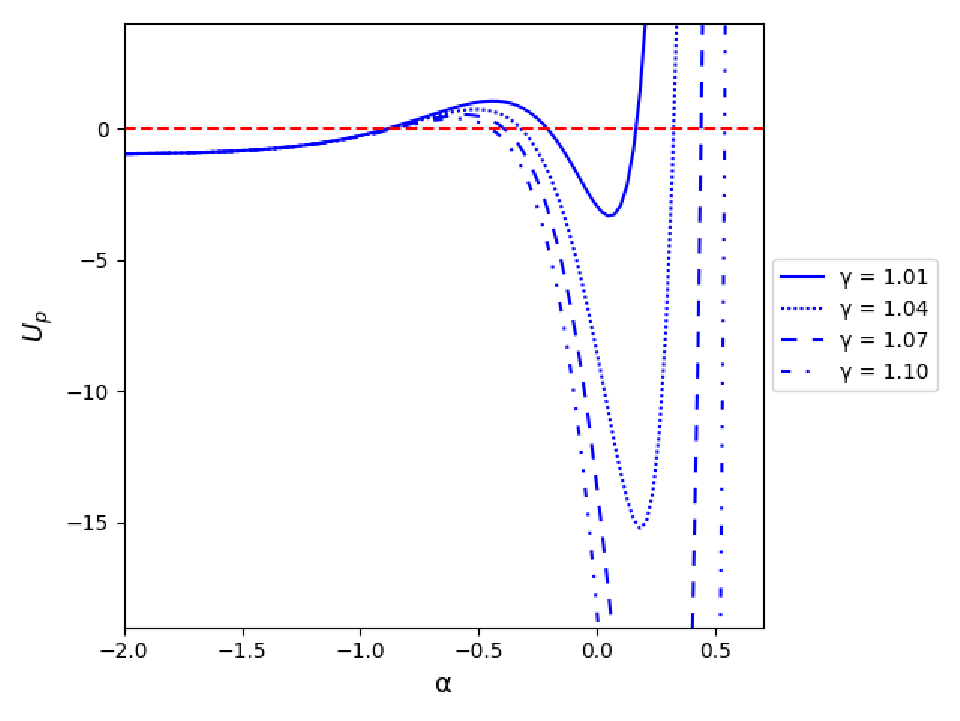}
  \caption{The potential (\ref{Up}) for several values of $\gamma$, with fixed
  $\beta=0.1$ and $p=1$ as well as $k=1$. \label{f:Up}}
\end{center}
\end{figure}

As shown in Figure~\ref{f:Up}, the contribution from the scalar field can now
be seen as opening up a classically allowed region around $\alpha\to-\infty$,
making it possible for the universe to tunnel from the trapped region, formed
by the other contributions, to a big-bang singularity. On the other side of
the $\alpha$-axis, expansion to infinite size is prevented by the steep
positive potential contribution proportional to $e^{6\alpha}$, which is
implied by the negative cosmological constant. The density contribution
$\sigma/a$ provides a negative contribution to the potential $U_p(\alpha)$
that forms a trapped region at intermediate values of $\alpha$, separated from
the asymptotic free region around $\alpha\to-\infty$ by a positive barrier
implied by the curvature term. For the intermediate negative contribution to
form a trapped region, it must be dominant between the two regions implied by
the contributions from spatial curvature and the cosmological constant,
respectively. In the original Friedmann equation, the corresponding energy
density must therefore follow a behavior between the power laws $a^{-2}$ of
the curvature term and $a^0$ of the cosmological constant. This requirement
explains the non-standard $a$-dependence of the energy density $\sigma/a$.

The quasiclassical dynamics of a given classical system in canonical form is
obtained by viewing variables such as $\alpha$ and $p_{\alpha}$ as
expectation values of the corresponding operators, taken in an evolving
quantum state. Any non-harmonic potential then implies that these variables
couple to fluctuations, correlations, and higher moments, implying dynamics in
a higher-dimensional configuration space. Coupling terms can be derived by
equipping moments with a Poisson bracket and inserting them in the expectation
value of the Hamilton operator of the system, taken in the same state in which
the moments are computed \cite{EffAc,Karpacz}. In general, the usual central
moments do not immediately appear in canonically conjugate form, but suitable
canonical pairs exist locally thanks to the Darboux theorem. Such canonical
variables have been derived for moments up to fourth order
\cite{Bosonize,EffPotRealize}.

For second-order moments, as a first approximation, canonical moment variables
for a single pair of degrees of freedom, such as $(\alpha,p_{\alpha})$ here,
have been known for some time, discovered independently in a variety of fields
\cite{VariationalEffAc,GaussianDyn,EnvQuantumChaos,QHDTunneling,CQC,CQCFieldsHom}:
There is an independent canonical pair $(s,p_s)$ that describes second-order
moments according to
\begin{eqnarray}
  \Delta(\alpha^2) &=& s^2 \label{s}\\
  \Delta(\alpha p_{\alpha}) &=& sp_s\\
  \Delta(p_{\alpha}^2) &=& p_s^2+\frac{U}{s^2}\,, \label{ps}
\end{eqnarray}
where $U$ is a constant bounded from below by $U\geq \hbar^2/4$ by the
uncertainty relation. We are using a general notation for moments
\begin{equation}
  \Delta(A^aB^b)=\left\langle(\hat{A}-\langle\hat{A}\rangle)^a(\hat{B}-\langle\hat{B}\rangle)^b\right\rangle_{\rm
  symm}
\end{equation}
in completely symmetric (or Weyl) ordering. According to this notation, the
two variances for a single canonical pair are
$(\Delta\alpha)^2=\Delta(\alpha^2)$ and $(\Delta
p_{\alpha})^2=\Delta(p_{\alpha}^2)$ and the covariance is $\Delta(\alpha p_{\alpha})$.

If the classical Hamiltonian is $H(\alpha,p_{\alpha})$, the new canonical
variables for second-order moments can be introduced in an effective
Hamiltonian by performing a Taylor expansion of
$H(\alpha+\delta\alpha, p_{\alpha}+\delta p_{\alpha})$ around a generic pair
$(\alpha,p_{\alpha})$ and replacing terms quadratic in $\delta \alpha$ and
$\delta p_{\alpha}$ with the moments (\ref{s})--(\ref{ps}).  The effective
energy expression, derived from (\ref{constraint}) in the
cosmological model, then reads
\begin{equation} \label{Usecond}
  0=p_{\alpha}^2+p_s^2+ \frac{U}{s^2}+ U_p(\alpha)+ \frac{1}{2} U_p''(\alpha)
  s^2
\end{equation}
where $U_p(\alpha)$ is given in (\ref{Up}).  (The classical equation
(\ref{constraint}) is a Hamiltonian constraint, which in a quantization is
turned into a constraint operator that annihilates physical states. This
condition restricts not only expectation values of basic operators by an
equation approximated by (\ref{Usecond}) semiclassically, but also
fluctuations and higher moments of the state
\cite{EffCons,EffConsRel,EffConsComp}. Since the constraint does
not depend on $\phi$ but only on its momentum, we can assume that moment
constraints are solved by using restricted values for $\phi$-moments.
The latter do not appear in the effective constraint and we do not need
specific solutions for them.)

While the momentum dependence of (\ref{constraint}) is quadratic and does not
imply higher-order terms in the Taylor expansion, the potential is not
harmonic. We therefore ignore certain quantum corrections in a truncation to
second order in $s$, given by (\ref{Usecond}). Tunneling, a process during
which a wave packet splits up into at least two smaller packets, is likely to
depend on moments of order higher than two. We should therefore amend
(\ref{Usecond}) by suitable higher-order terms, while keeping the system
sufficiently simple for an initial analysis. In particular, higher-order
moments, in a canonical formulation, describe degrees of freedom
independent of both $(\alpha,p_{\alpha})$ and $(s,p_s)$, and therefore lead to
configuration spaces of large dimensions if they are included in complete
form.

As an approximation, it is possible to include some higher-moment
effects without higher dimensions by making an ansatz for the possible
behavior of moments on the quantum degree of freedom $(s,p_s)$ already
introduced for second order. Dimensional arguments suggest the power-law form
$\Delta(\alpha^n)\propto s^n$ for an $\alpha$-moment of order $n$. If this
form were realized exactly with coefficient one, the Taylor expansion of the
effective potential could be summed up analytically:
\begin{equation} \label{Taylor}
  0=p_{\alpha}^2+ p_s^2+ \frac{U}{s^2}+
  U_p(\alpha)+ \sum_{n=2}^{\infty}\frac{U_p^{(n)}}{n!}\Delta(\alpha^n)
  =p_{\alpha}^2+p_s^2+\frac{U}{s^2}
   +\frac{1}{2}\left(U_p(\alpha+s)+U_p(\alpha-s)\right)\,.
\end{equation}
As in \cite{TunnelingUniverse}, following
\cite{QuantumHiggsInflation,EffPotInflation}, we bring the fourth-order term
closer to Gaussian form, where $\Delta(\alpha^4)=3s^4$ rather than $s^4$, by
adding the final term in
\begin{equation} \label{UAllFour}
   0=p_{\alpha}^2+p_s^2+\frac{U}{s^2}
   +\frac{1}{2}\left(U_p(\alpha+s)+U_p(\alpha-s)\right)+ \frac{1}{12} U_p^{''''}(\alpha)s^4\,.
\end{equation}
Our analysis of chaos will use mainly the small-$s$ dynamics in the trapped
region before much tunneling happens, in which case the quasiclassical approximation is
expected to be reliable.

\begin{figure}
\begin{center}
  \includegraphics[width=16cm]{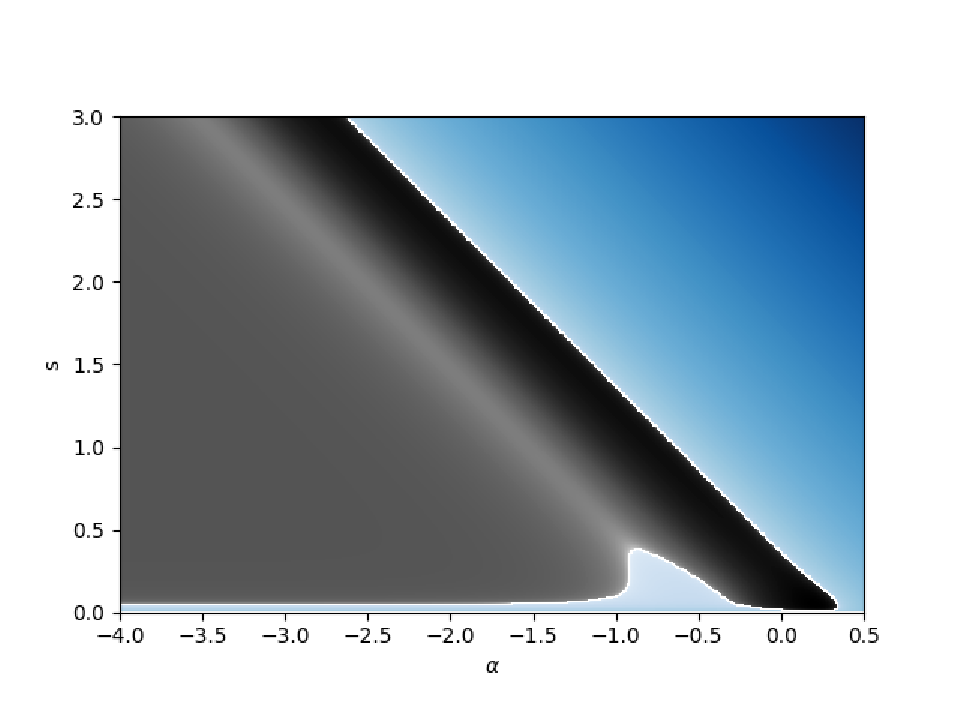}
  \caption{Contour plot of the quasiclassical potential in
    (\ref{Taylor}). Negative values of the potential are shown in shades of
    gray, while positive values are in blue. The black region crossing the
    figure in a diagonal way is the deep channel that extends the trapped
    region into the $s$-dimension. The thinner light-gray strip that bounds
    the channel on its left flank is an elevation in the region of negative
    potential implied by the classical barrier. For a complete description of
    tunneling, trajectories in the quasiclassical model must cross the
    elevation once in the past and in the future of being
    trapped.  \label{f:Log}}
\end{center}
\end{figure}

\section{Tunneling and chaos}

\begin{figure}
\begin{center}
  \includegraphics[width=16cm]{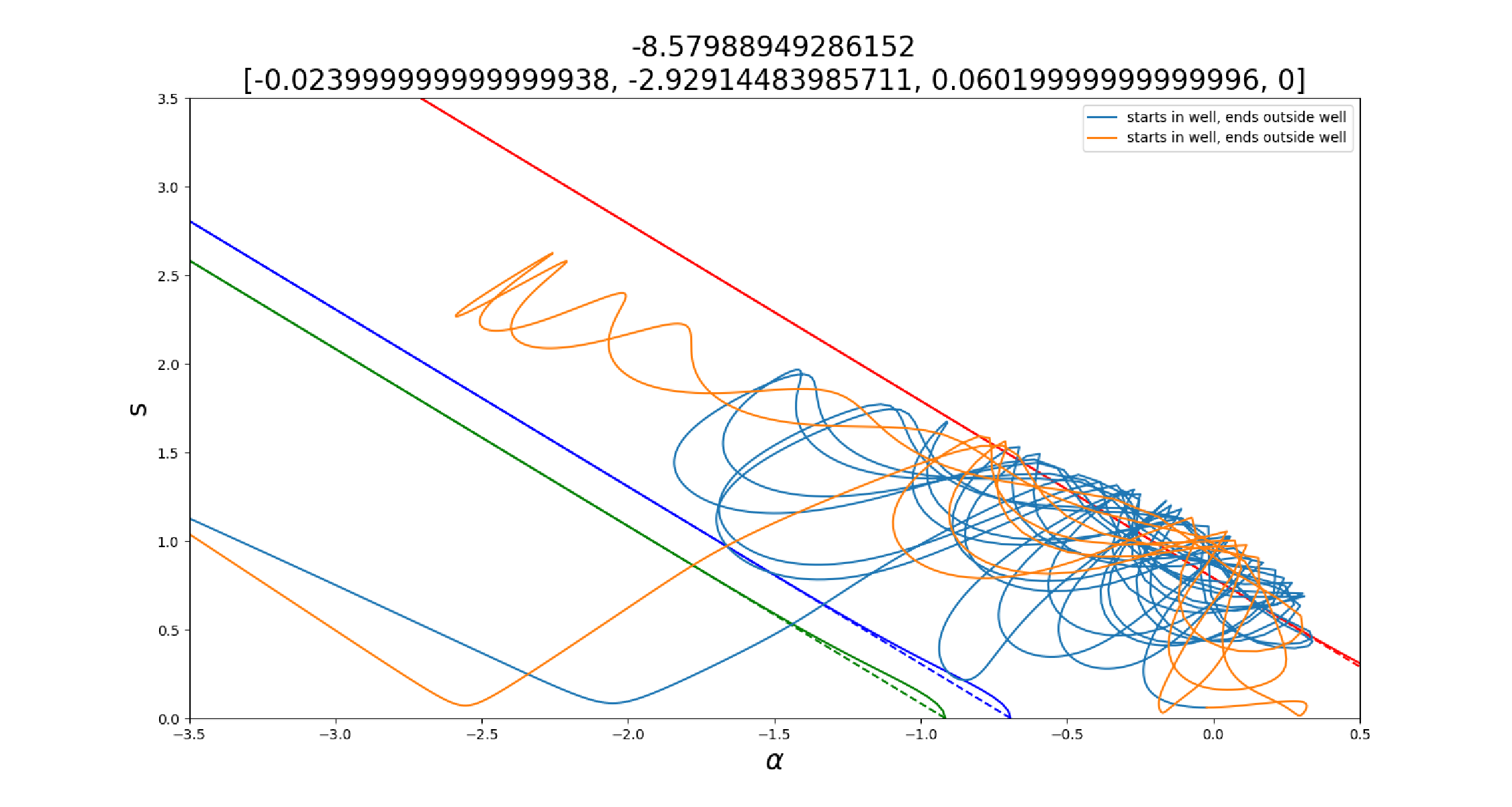}
  \caption{Example of a trajectory that tunnels in the past and in the
    future. The number in the top line indicates the initial potential, while
    initial phase-space values are given in the second line in the form
    $[\alpha,p_{\alpha},s,p_s]$.  \label{f:Free}}
\end{center}
\end{figure}

Tunneling is quasiclassically described by motion in the $(\alpha,s)$-plane,
relying on several characteristic features of the effective potential in
(\ref{UAllFour}); see Figure~\ref{f:Log}. By the addition and subtraction of
$s$ in the terms resulting from (\ref{Taylor}), the classically trapped region
is extended to a diagonal channel in the $s$-direction. The channel is bounded
by a finite wall to the left and a steep increasing wall to the right. Since
the wall on the left has a height lower than the classical barrier for
sufficiently large $s$, tunneling is possible when a quasiclassical trajectory
crosses over the wall into the unbounded region to the left, where it will
then continue almost freely except for one possible reflection at the $U/s^2$
potential; see Fig.~\ref{f:Free}.

\begin{figure}
\begin{center}
  \includegraphics[width=16cm]{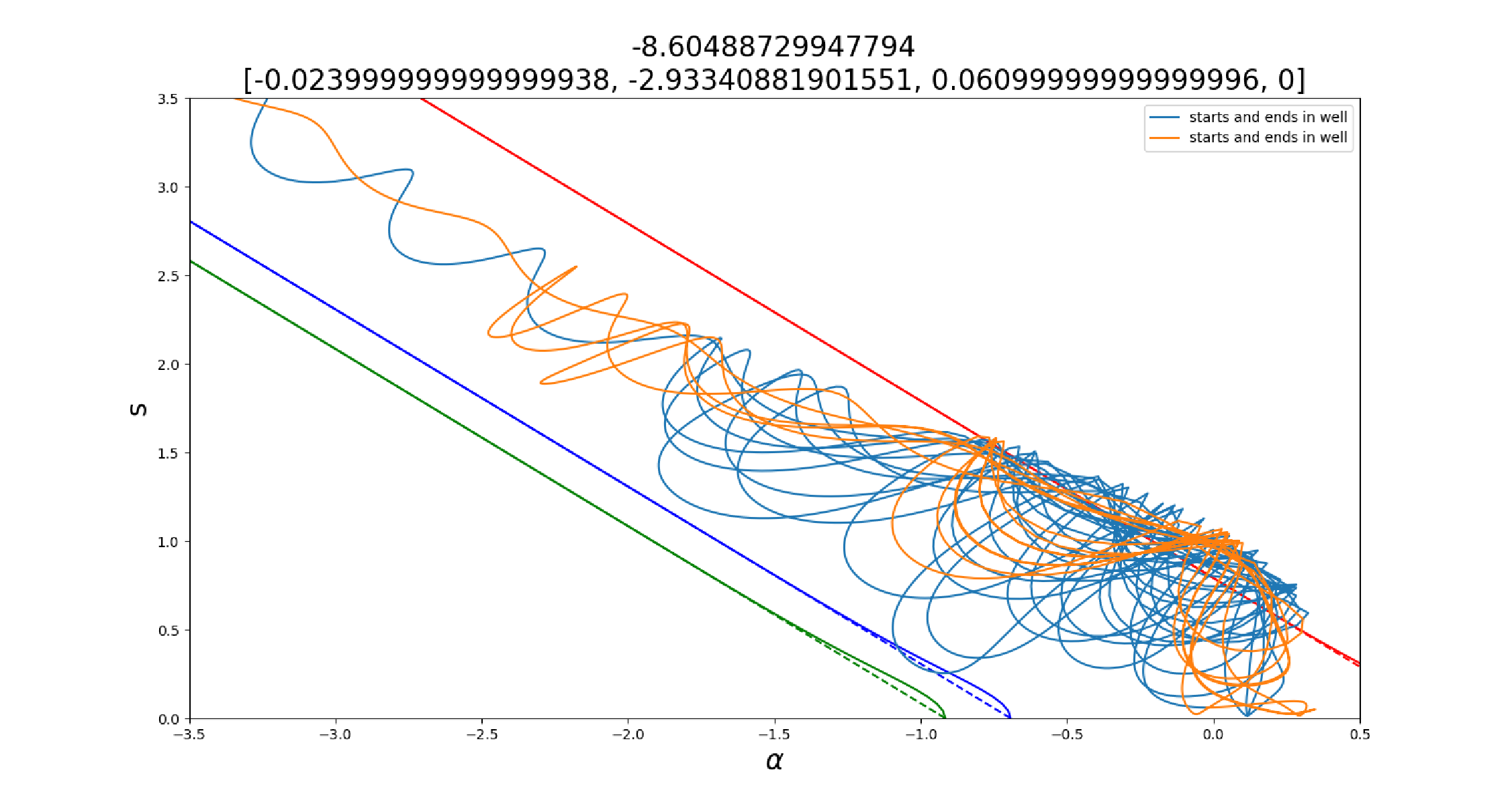}
  \caption{Example of a trajectory that never tunnels. The number in the top
    line indicates the initial potential, while initial phase-space values are
    given in the second line in the form
    $[\alpha,p_{\alpha},s,p_s]$.  \label{f:Trapped}}
\end{center}
\end{figure}

However, since the quasiclassical model does not capture all features of
tunneling, there are also trajectories that never cross over the wall even if
their energy would be sufficient for full quantum mechanical tunneling; see
Fig.~\ref{f:Trapped}. Such trajectories are stuck in the channel and keep
bouncing between the two walls, moving to ever larger $s$. Here, the
quasiclassical approximation will eventually break down.

\begin{figure}
\begin{center}
  \includegraphics[width=16cm]{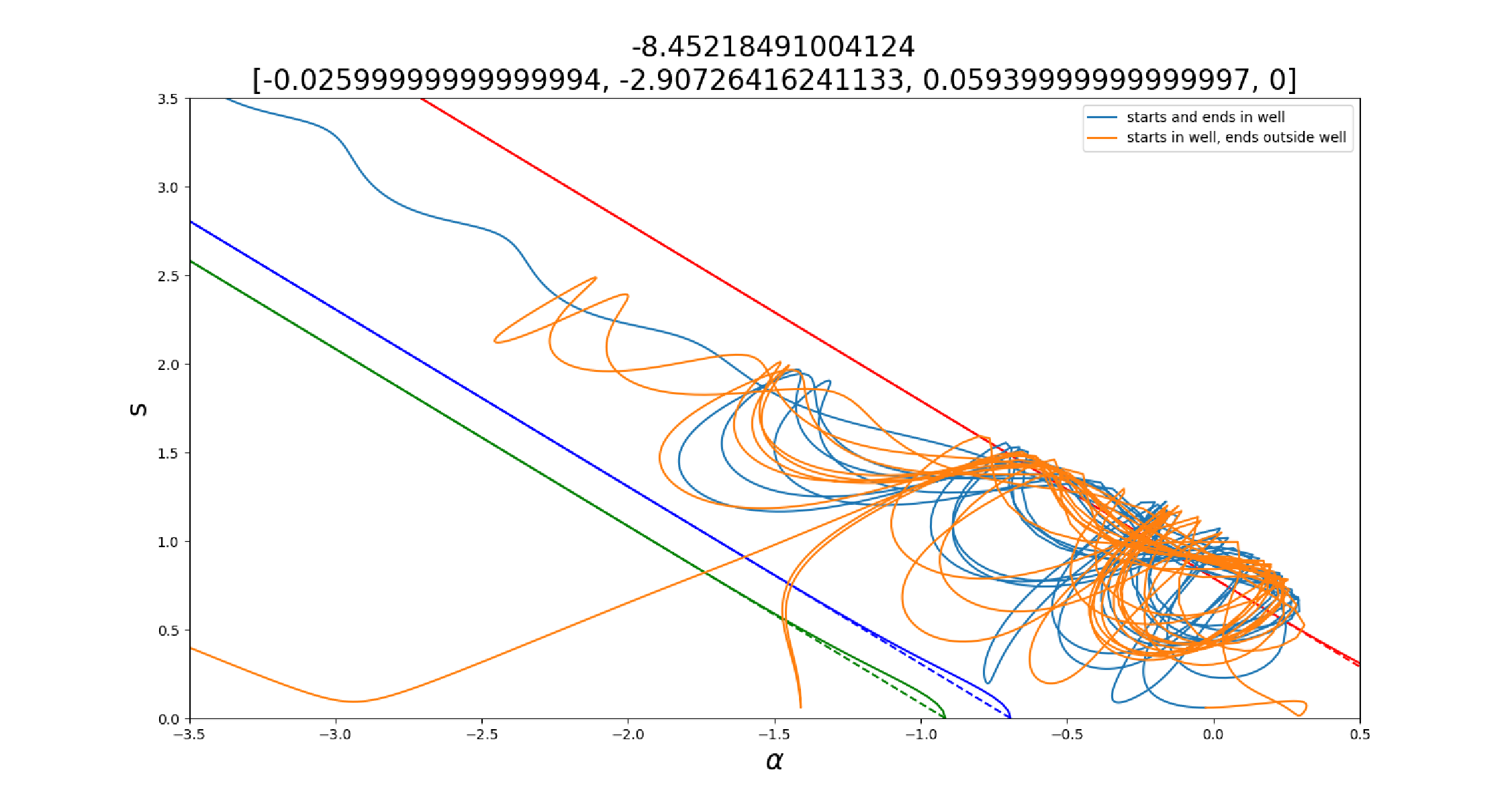}
  \caption{Example of a trajectory that tunnels only once in the past or
    future. The number in the top line indicates the initial potential, while
    initial phase-space values are given in the second line in the form
    $[\alpha,p_{\alpha},s,p_s]$.  \label{f:Partial}}
\end{center}
\end{figure}

Trajectories stuck in the channel therefore do not describe correct features
of tunneling, but their presence allows us to draw an important distinction
between two types of trajectories: Those that tunnel correctly by moving over
the wall on the left, and those that get stuck in the channel. The latter can
be split into subcases of trajectories getting stuck only to one side in time
(future or past; see Fig.~\ref{f:Partial}) or to both sides. Physically, any
trajectory that gets stuck corresponds to a wave function for which higher
moments are relevant, while trajectories that do not get stuck have a
tunneling process well described by lower moments.

\begin{figure}
\begin{center}
  \includegraphics[width=12cm]{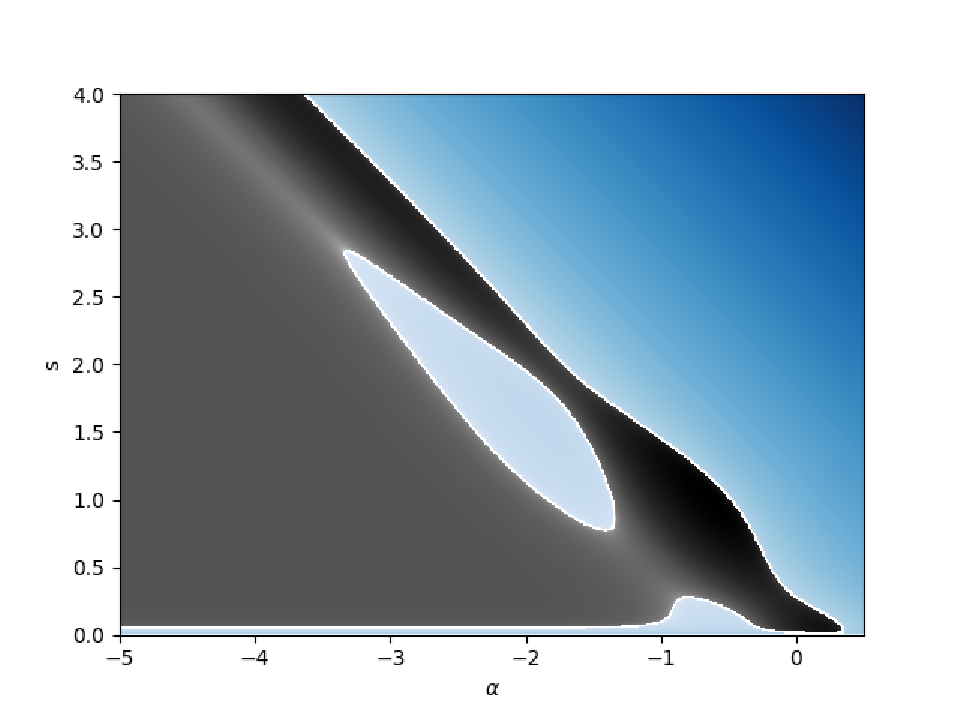}
  \caption{Contour plot of the modified quasiclassical potential
    (\ref{UAllFour}) with new features in the small-$s$ part of the
    channel. While keeping a door open to exit the channel at small $s$, the
    channel itself narrows such that it is less likely for quasiclassical
    trajectories to follow it to large $s$. At the same time, the new features
    introduce additional convex or defocusing contributions to the channel
    walls at small $s$, which are expected to enhance
    chaos.  \label{f:ChannelFour}}
\end{center}
\end{figure}

A detailed analysis of trajectories, given in \cite{TunnelingUniverse}, showed
that the potential (\ref{UAllFour}) does not reliably describe tunneling
because the uniform nature of the channel, seen in Figure~\ref{f:Log}, makes
it much more likely for trajectories to follow the channel, rather than
crossing the wall to the left. The fourth-order modification in
(\ref{UAllFour}), motivated by a more Gaussian behavior of states, was found
to improve the tunneling description by quasiclassical trajectories. With this
modification, the channel acquires new features at small $s$, shown in
Figure~\ref{f:ChannelFour}, that can help to turn trajectories toward the
channel wall.

\begin{figure}
\begin{center}
  \includegraphics[width=12cm]{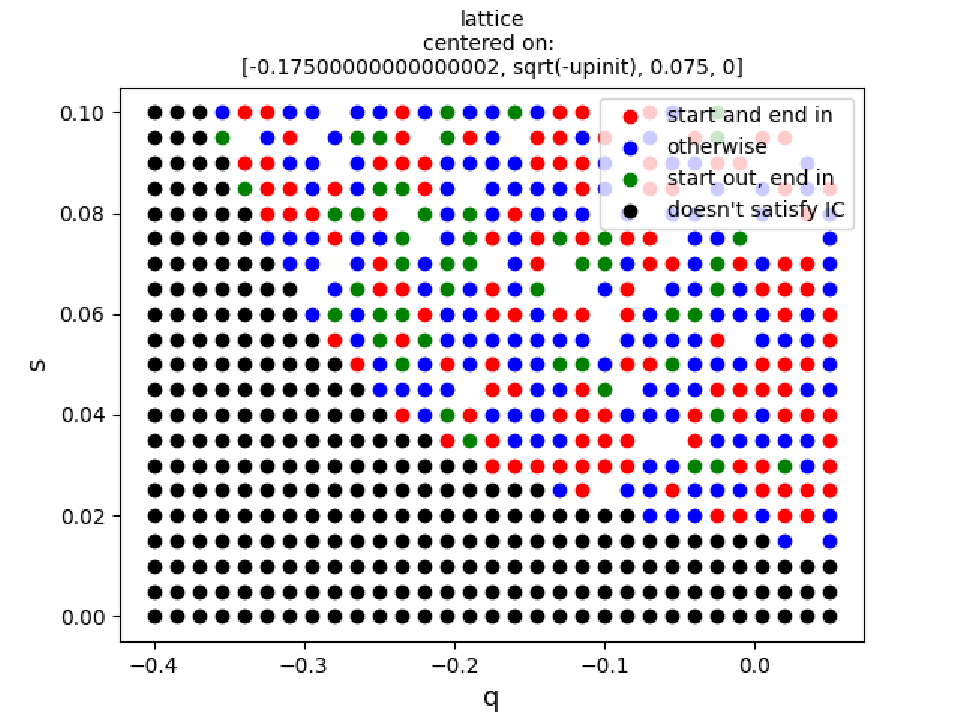}
  \caption{An initial lattice of outcomes, depending on the initial values of
    $q=\alpha$ and $s$. Within this lattice, initial points are marked with
    colors depending on the long-term outcome of trajectories starting there,
    all with the same initial momenta $p_{\alpha}=0.075$ and $p_s=0$. Empty
    (or white) circles indicate inconclusive outcomes because the running time
    was too long. The black region is outside the channel walls and does not
    lead to tunneling.  \label{f:lattice20}}
\end{center}
\end{figure}

There were two indications for chaos in this dynamics found in
\cite{TunnelingUniverse}: A sensitive dependence of long-term outcomes of
trajectories on their initial values in the classically trapped region; and
the shape of the confining walls around the classically trapped region,
extended to the $(\alpha,s)$-plane. Since the latter have convex or
defocussing contributions, especially with the fourth-order modification as
seen in Figure~\ref{f:ChannelFour}, mathematical arguments from dynamical
billiard systems may be used to infer the possibility of chaotic features
\cite{Sinai}, as done also in other cosmological models
\cite{Billiards}. However, the walls are not completely convex. A dedicated
analysis is therefore required to determine properties of chaos
\cite{FocusingChaos}. We do so now by numerical computations of the fractal
dimension of sets of initial values in the bottom part of the channel that
give rise to the same long-term outcome of tunneling trajectories. The
sensitivity to the choice of initial values is illustrated by
Figure~\ref{f:lattice20}.

\begin{figure}
\begin{center}
  \includegraphics[width=12cm]{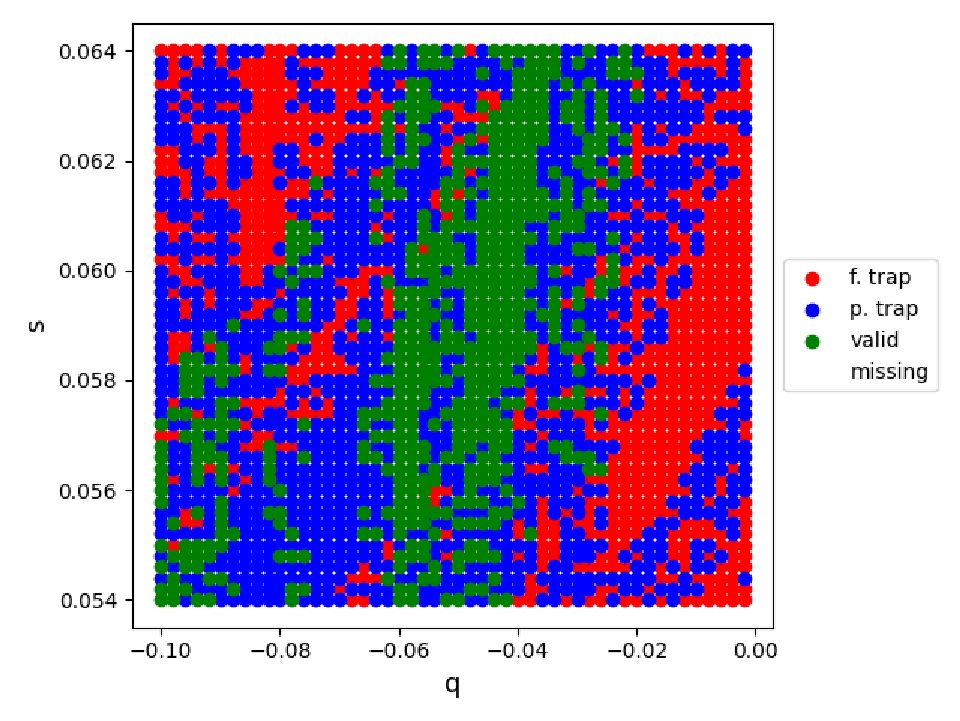}
  \caption{A finer sublattice of Fig.~\ref{f:lattice20}, revealing new
    structures of independent outcomes. 
    \label{f:lattice2500}}
\end{center}
\end{figure}

\begin{figure}
\begin{center}
  \includegraphics[width=12cm]{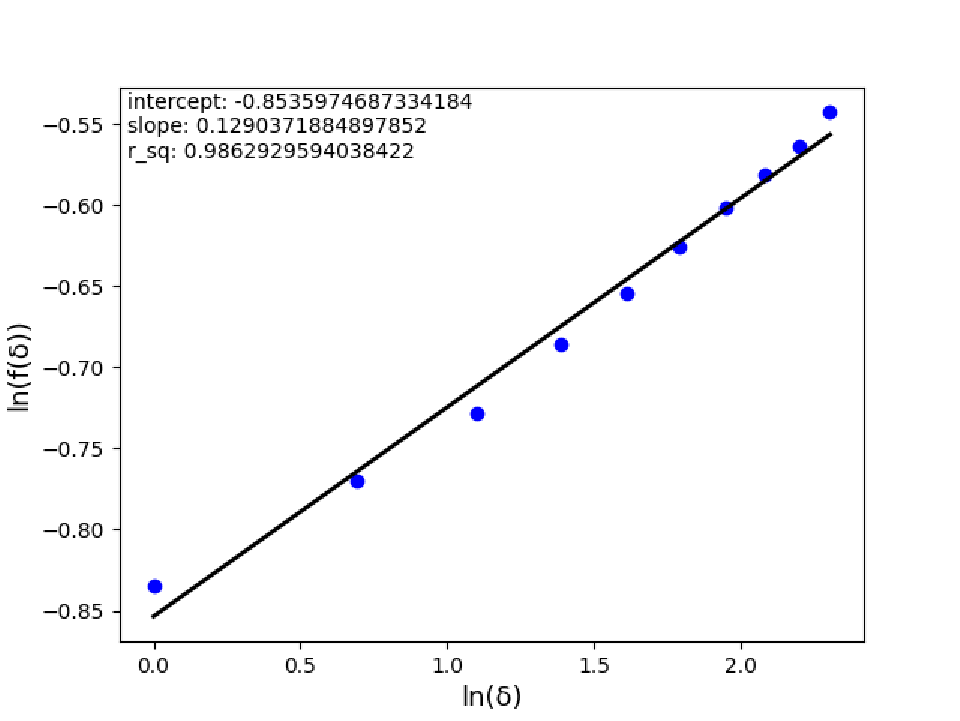}
  \caption{Linear fit to logarithm of (\ref{fdelta}). The slope of the fit is
    our result for $\epsilon$.
  \label{f:FractalDim}}
\end{center}
\end{figure}

The procedure for calculating the fractal dimension of the model and
demonstrating chaos involved generating a $50\times 50$ lattice of distinct
tunneling results, illustrated in Figure~\ref{f:lattice2500}. The points in
the sample region were classified as fully trapped, partially trapped, or
untrapped based on the final state of the model. The lattice was analyzed
using the uncertainty exponent analysis for sensitive fractal boundaries
demonstrated in \cite{BasinChaos}, a function which scales as a power of the
radius (or the size of the basin boundary),
\begin{equation}\label{fdelta}
  f(\delta)\sim \delta^{\epsilon}
\end{equation}
where $\epsilon$ is the uncertainty coefficient. Points are taken within a
radius $\delta$, centered sequentially on each lattice point. The fraction of
points within $\delta$ demonstrating different final states from the initial
point was calculated for each lattice element, as in
Figure~\ref{f:FractalDim}. Following \cite{BasinChaos}, systems where
$\epsilon=1$ are not chaotic --- no uncertainty appears when varying initial
conditions. The range of $\epsilon$ for chaotic systems is given by
$0\leq\epsilon<1$. Lower values of $\epsilon$ indicate that the system is more
chaotic. As emphasized in \cite{Farey}, measuring chaos using this basin
method \cite{BasinChaos} is more suitable for relativistic or time
reparameterization invariant systems because it depends only on the final
outcomes of trajectories and not on their parameterization by time, unlike for
instance the computation of Lyapunov exponents. Reparameterization of the time
variable will not alter the final states indicated on the lattice. As also
used recently in \cite{QuasiClassChaos}, the method can easily be generalized
to quasiclassical descriptions of quantum systems.

Furthermore, $\epsilon$ satisfies \cite{BasinChaos}
\begin{equation}
  \epsilon=D-D_0
\end{equation}
where $D$ is the dimension of the phase space and $D_0$ indicates the
dimension of the boundary which divides the regions with different
outcomes. In chaotic systems, $D_0$ assumes a non-integer value which means
that the boundaries are fractal.  Computing the value of $\epsilon$ for the
model produced Figure~\ref{f:FractalDim}, using values $1\leq\delta\leq 10$.
A linear fit of the data in a double logarithmic plot revealed the slope
$\epsilon=0.129$, see Figure~\ref{f:FractalDim}, and is thus a reliable indicator
of chaos.

\section{Conclusions}

Our results demonstrate that the quasiclassical dynamics of the oscillating
universe model studied here is chaotic. The classical system has a
1-dimensional configuration space and therefore cannot have chaos, but
quantization implies additional independent degrees of freedom such as the
fluctuation parameter $s$ in our analysis. The appearance of chaos here is
therefore distinct from the traditional notion of quantum chaos, which is
usually analyzed in situations in which the classical system is already
chaotic.

Our model is closer to discussions of chaos in Bohmian quantum mechanics, such
as \cite{BohmianChaos}, in which the quantum potential plays the role of our
quasiclassical potential. Compared with Bohmian quantum mechanics, our
analysis is held completely at a phase-space level as in classical
mechanics. Quantum effects are described by moments in a canonical
parameterization, implying new configuration variables and momenta. The full
wave function is approximated and ultimately replaced by the moments and does
not play an intermediary role, for instance as the wave function of Bohmian
quantum mechanics used to compute the quantum potential. Standard quantitative
methods to analyze chaos can therefore be applied directly.

A qualitative argument, using the convex nature of some portions of the
potential walls bounding the quasiclassical trapped region, demonstrate a
relationship between chaos and detailed properties of the quantum state. In
particular, our quantitative results about chaos refer to a quasiclassical
potential with fourth-order moments of Gaussian form, which compared with
other choices of moments leads to more convex walls as seen in
Figure~\ref{f:ChannelFour}. Our results therefore suggest that details of
quantum states and properties of their quantum information may have direct
implications on important features of the early-universe dynamics.

\acknowledgments

We are grateful to Sara Fern\'{a}ndez Uria for bringing our attention to the
basin method and for discussions about it. This work was supported in part by
NSF grant PHY-2206591.

%\bibliographystyle{JHEP}
%\bibliography{../Bib/QuantGra,../Bib/Tunneling}

\begin{thebibliography}{10}

\bibitem{tunneling}
A.~Vilenkin, {\it Quantum creation of universes},  {\em Phys.\ Rev.\ D} {\bf
  30} (1984) 509--511.

\bibitem{OscillatingFriedmann}
M.~P. D\c{a}browski, {\it Oscillating friedman cosmology},  {\em Ann.\ Phys.}
  {\bf 248} (1996) 199--219, [\href{http://xxx.lanl.gov/abs/gr-qc/9503017}{{\tt
  gr-qc/9503017}}].

\bibitem{OscillatingSimple}
P.~W. Graham, B.~Horn, S.~Kachru, S.~Rajendran, and G.~Torroba, {\it A simple
  harmonic universe},  {\em JHEP} {\bf 02} (2014) 029,
  [\href{http://xxx.lanl.gov/abs/1109.0282}{{\tt 1109.0282}}].

\bibitem{OscillatingTunnel}
M.~P. D\c{a}browski and A.~L. Larsen, {\it Quantum tunneling effect in
  oscillating friedmann cosmology},  {\em Phys.\ Rev.\ D} {\bf 52} (1995)
  3424--3431, [\href{http://xxx.lanl.gov/abs/gr-qc/9504}{{\tt gr-qc/9504}}].

\bibitem{OscillatingCollapse}
A.~T. Mithani and A.~Vilenkin, {\it Collapse of simple harmonic universe},
  {\em JCAP} {\bf 01} (2012) 028,
  [\href{http://xxx.lanl.gov/abs/1110.4096}{{\tt 1110.4096}}].

\bibitem{OscillatingRate}
A.~T. Mithani and A.~Vilenkin, {\it Tunneling decay rate in quantum cosmology},
   {\em Phys.\ Rev.\ D} {\bf 91} (2015) 23511,
  [\href{http://xxx.lanl.gov/abs/1503.00400}{{\tt 1503.00400}}].

\bibitem{TunnelingUniverse}
M.~Bojowald and P.~Petersen, {\it Tunneling dynamics of an oscillating universe
  model},  {\em JCAP} {\bf 05} (2022) 007,
  [\href{http://xxx.lanl.gov/abs/2110.09491}{{\tt 2110.09491}}].

\bibitem{ChaosGR}
J.~D. Barrow, {\it Chaotic behaviour in general relativity},  {\em Phys.\ Rep.}
  {\bf 85} (1982) 1--49.

\bibitem{Farey}
N.~J. Cornish and J.~J. Levin, {\it The mixmaster universe: A chaotic farey
  tale},  {\em Phys.\ Rev.\ D} {\bf 55} (1997) 7489.

\bibitem{ChaosRel}
A.~E. Motter, {\it Relativistic chaos is coordinate invariant},  {\em Phys.\
  Rev.\ Lett.} {\bf 91} (2003) 231101.

\bibitem{Billiards}
T.~Damour, M.~Henneaux, and H.~Nicolai, {\it Cosmological billiards},  {\em
  Class.\ Quantum Grav.} {\bf 20} (2003) R145--R200,
  [\href{http://xxx.lanl.gov/abs/hep-th/0212256}{{\tt hep-th/0212256}}].

\bibitem{BKL}
V.~A. Belinskii, I.~M. Khalatnikov, and E.~M. Lifschitz, {\it A general
  solution of the einstein equations with a time singularity},  {\em Adv.\
  Phys.} {\bf 31} (1982) 639--667.

\bibitem{Isotropize}
C.~W. Misner, {\it The isotropy of the universe},  {\em Astrophys.\ J.} {\bf
  151} (1968) 431--457.

\bibitem{Infrared}
M.~Bojowald, {\it The bkl scenario, infrared renormalization, and quantum
  cosmology},  {\em JCAP} {\bf 01} (2019) 026,
  [\href{http://xxx.lanl.gov/abs/1810.00238}{{\tt 1810.00238}}].

\bibitem{Foundations}
M.~Bojowald, {\em Foundations of Quantum Cosmology}.
\newblock IOP Publishing, London, UK, 2020.

\bibitem{EffAc}
M.~Bojowald and A.~Skirzewski, {\it Effective equations of motion for quantum
  systems},  {\em Rev.\ Math.\ Phys.} {\bf 18} (2006) 713--745,
  [\href{http://xxx.lanl.gov/abs/math-ph/0511043}{{\tt math-ph/0511043}}].

\bibitem{Karpacz}
M.~Bojowald and A.~Skirzewski, {\it Quantum gravity and higher curvature
  actions},  {\em Int.\ J.\ Geom.\ Meth.\ Mod.\ Phys.} {\bf 4} (2007) 25--52,
  [\href{http://xxx.lanl.gov/abs/hep-th/0606232}{{\tt hep-th/0606232}}].
  Proceedings of ``Current Mathematical Topics in Gravitation and Cosmology''
  (42nd Karpacz Winter School of Theoretical Physics), Ed.\ Borowiec, A.\ and
  Francaviglia, M.

\bibitem{Bosonize}
B.~Bayta\c{s}, M.~Bojowald, and S.~Crowe, {\it Faithful realizations of
  semiclassical truncations},  {\em Ann.\ Phys.} {\bf 420} (2020) 168247,
  [\href{http://xxx.lanl.gov/abs/1810.12127}{{\tt 1810.12127}}].

\bibitem{EffPotRealize}
B.~Bayta\c{s}, M.~Bojowald, and S.~Crowe, {\it Effective potentials from
  canonical realizations of semiclassical truncations},  {\em Phys.\ Rev.\ A}
  {\bf 99} (2019) 042114, [\href{http://xxx.lanl.gov/abs/1811.00505}{{\tt
  1811.00505}}].

\bibitem{VariationalEffAc}
R.~Jackiw and A.~Kerman, {\it Time dependent variational principle and the
  effective action},  {\em Phys.\ Lett.\ A} {\bf 71} (1979) 158--162.

\bibitem{GaussianDyn}
F.~Arickx, J.~Broeckhove, W.~Coene, and P.~van Leuven, {\it Gaussian
  wave-packet dynamics},  {\em Int.\ J.\ Quant.\ Chem.: Quant.\ Chem.\ Symp.}
  {\bf 20} (1986) 471--481.

\bibitem{EnvQuantumChaos}
R.~A. Jalabert and H.~M. Pastawski, {\it Environment-independent decoherence
  rate in classically chaotic systems},  {\em Phys.\ Rev.\ Lett.} {\bf 86}
  (2001) 2490--2493.

\bibitem{QHDTunneling}
O.~Prezhdo, {\it Quantized hamiltonian dynamics},  {\em Theor.\ Chem.\ Acc.}
  {\bf 116} (2006) 206.

\bibitem{CQC}
T.~Vachaspati and G.~Zahariade, {\it A classical-quantum correspondence and
  backreaction},  {\em Phys.\ Rev.\ D} {\bf 98} (2018) 065002,
  [\href{http://xxx.lanl.gov/abs/1806.05196}{{\tt 1806.05196}}].

\bibitem{CQCFieldsHom}
M.~Mukhopadhyay and T.~Vachaspati, {\it Rolling with quantum fields},
  \href{http://xxx.lanl.gov/abs/1907.03762}{{\tt 1907.03762}}.

\bibitem{EffCons}
M.~Bojowald, B.~Sandh\"ofer, A.~Skirzewski, and A.~Tsobanjan, {\it Effective
  constraints for quantum systems},  {\em Rev.\ Math.\ Phys.} {\bf 21} (2009)
  111--154, [\href{http://xxx.lanl.gov/abs/0804.3365}{{\tt 0804.3365}}].

\bibitem{EffConsRel}
M.~Bojowald and A.~Tsobanjan, {\it Effective constraints for relativistic
  quantum systems},  {\em Phys.\ Rev.\ D} {\bf 80} (2009) 125008,
  [\href{http://xxx.lanl.gov/abs/0906.1772}{{\tt 0906.1772}}].

\bibitem{EffConsComp}
M.~Bojowald and A.~Tsobanjan, {\it Effective constraints and physical coherent
  states in quantum cosmology: A numerical comparison},  {\em Class.\ Quantum
  Grav.} {\bf 27} (2010) 145004, [\href{http://xxx.lanl.gov/abs/0911.4950}{{\tt
  0911.4950}}].

\bibitem{QuantumHiggsInflation}
M.~Bojowald, S.~Brahma, S.~Crowe, D.~Ding, and J.~McCracken, {\it Quantum higgs
  inflation},  {\em Phys.\ Lett.\ B} {\bf 816} (2021) 136193,
  [\href{http://xxx.lanl.gov/abs/2011.02355}{{\tt 2011.02355}}].

\bibitem{EffPotInflation}
M.~Bojowald, S.~Brahma, S.~Crowe, D.~Ding, and J.~McCracken, {\it Multi-field
  inflation from single-field models},  {\em JCAP} {\bf 08} (2021) 047,
  [\href{http://xxx.lanl.gov/abs/2011.02843}{{\tt 2011.02843}}].

\bibitem{Sinai}
Y.~G. Sinai, {\it Dynamical systems with elastic reflections. ergodic
  properties of dispersing billiards.},  {\em Russian Mathematical Surveys}
  {\bf 25} (1970) 137--189.

\bibitem{FocusingChaos}
L.~A. Bunimovich, {\it On billiards close to dispersing},  {\em Mathematical
  USSR Sbornik} {\bf 95} (1974) 49--73.

\bibitem{BasinChaos}
S.~W. McDonald, C.~Grebogi, E.~Ott, and J.~A. Yorke, {\it Fractal basin
  boundaries},  {\em Physica D: Nonlinear Phenomena} {\bf 17} (1985) 125.

\bibitem{QuasiClassChaos}
M.~Bojowald, D.~Brizuela, P.~Calizaya~Cabrera, and S.~Uria, {\it The chaotic
  behavior of the bianchi ix model under the influence of quantum effects},
  \href{http://xxx.lanl.gov/abs/2307.00063}{{\tt 2307.00063}}.

\bibitem{BohmianChaos}
G.~Contopoulos and A.~C. Tzemos, {\it Chaos in bohmian quantum mechanics: A
  short review},  {\em Regul.\ Chaot.\ Dyn.} {\bf 25} (2020) 476--495,
  [\href{http://xxx.lanl.gov/abs/2009.05867}{{\tt 2009.05867}}].

\end{thebibliography}

\providecommand{\href}[2]{#2}\begingroup\raggedright\endgroup

\end{document}